\documentclass{PoS}

\usepackage{graphicx}

\def\be{\begin{eqnarray*}}
\def\ee{\end{eqnarray*}}
\def\bea{\begin{eqnarray}}
\def\eea{\end{eqnarray}}

\def\ket#1{\left|#1\right>}

\title{Projection of the low-lying eigenmodes of the overlap Dirac operator in lattice QCD}

\ShortTitle{Projection of low-modes in lattice QCD}

\author{Ting-Wai Chiu$^{1,2}$, \speaker{Tung-Han Hsieh}$^3$
	(TWQCD Collaboration)\\
        $^1$ Department of physics, National Taiwan University, Taipei 10617, Taiwan\\
	$^2$ Center for Quantum Science and Engineering, National Taiwan University, \\ 
        \hspace{2mm} Taipei 10617, Taiwan\\
	$^3$ Research Center for Applied Sciences, Academia Sinica, Taipei 115, Taiwan
        }

\abstract{%
We outline our implementation of the adaptive thick-restart Lanczos algorithm ($a$-TRLan) for the  
projection of the low-lying eigenmodes of the overlap Dirac operator in lattice QCD, and compare 
the performances of our code and the widely used package ARPACK. 
}

\FullConference{2013 International Workshop on Computational Science
		and Engineering,\\
		14-17 October 2013, Taipei, Taiwan}

\begin{document}

\section{Introduction}

To compute all eigenmodes of a large sparse matrix is a very challenging problem. 
However, for most studies, only a small subset of the entire spectrum plays the dominant role. 
Thus it is feasible to use an iterative algorithm to {\em project} this subset of eigenmodes.
  
In lattice QCD, it is well-known that the low-lying
eigenmodes $(\lambda_i, q_i)$ of the Hermitian Dirac operator $H$
(where $H = \gamma_5 D$, and $D$ is the Dirac operator which satisfies
the $\gamma_5$-Hermiticity, $D^{\dagger} = \gamma_5 D\gamma_5$)
can serve as a basis to tackle many problems. 
For example, to compute the quark propagator $ x = D^{-1} b $, 
it amounts to solving the linear system $H^2 x = D^\dagger b \equiv c $ with the conjugate gradient algorithm (CG). 
Since the convergence rate of CG is proportional to the condition number 
$\lambda_{\mathrm{max}}^2/\lambda_{\mathrm{min}}^2$ (where
$\lambda_{\mathrm{max}} / \lambda_{\mathrm{min}}$ are the largest/smallest eigenvalues of $|H|$), 
the convergence can be accelerated with the low-mode preconditioning. 
With the projected low modes $(\lambda_i, q_i) $ of $ H^2 $, 
CG can be performed in the orthogonal complement
of the subspace spanned by the low modes, i.e.,  
\be
H^2 \tilde x = \left( I - \sum_i q_i q_i^{\dagger} \right) c,
\ee
and the final solution is  
\be
x = \tilde x + \sum_i \frac{1}{\lambda_i^2}q_i q_i^{\dagger} c.
\ee

Another application of the low-lying eigenmodes (of the Wilson-Dirac operator)
is to obtain a better approximation (chiral symmetry) of the overlap-Dirac operator 
$D_{\mathrm{ov}}$ for massless fermion \cite{Neuberger:1997fp} 
\bea
\label{Dovb}
D_{\mathrm{ov}} = m_0 \left[ 1 + \gamma_5\, \mathrm{sgn}(H_w) \right],  \hspace{4mm}
{\rm sgn}(H_w) = \frac{H_w}{\sqrt{H_w^2}}, 
\eea
where $H_w$ is the Hermitian Wilson-Dirac operator plus a negative parameter
$-m_0$ ($0 < m_0 < 2$). Since $(H_w^2)^{-1/2}$ cannot be evaluated
exactly, one often uses the {\em Zolotarev optimal rational approximation} \cite{Chiu:2002eh}. 
For a given order of the optimal rational polynomial $R_Z(H_w^2)$, it attains the
optimal approximation of $(H_w^2)^{-1/2}$ in the interval  
$[\lambda_{\mathrm{min}}^2, \lambda_{\mathrm{max}}^2]$ of the spectrum of $H_w^2$.
Moreover, for a given order, a smaller ratio $(\lambda_{\mathrm{max}}^2/\lambda_{\mathrm{min}}^2)$ gives 
a better approximation. Thus, with the projected low modes of $ H_w^2 $, 
the matrix-vector product ($ D_{\rm ov} v $) can be decomposed into
the low-mode and the high-mode parts as 
\bea
\label{Dovb_with_Hw_low-mode}
D_{\mathrm{ov}} v = m_0\biggl\{ 1 + \gamma_5
\sum_i q_i\,\mathrm{sgn}(\lambda_i) q_i^{\dagger}  +
\gamma_5\, H_w R_Z(H_w^2) \Bigl(I - \sum_i q_i q_i^{\dagger}\Bigr) \biggr\} v, 
\eea
which yields better chiral symmetry than those without low modes.  

The physical significances of the low-modes of the overlap Dirac operator are related to the 
topological quantum fluctuations of the QCD vacuum.
According to the Atiyah-Singer index theorem \cite{Atiyah:1968mp}, 
the numbers of zero modes ($n_\pm$) of the overlap Dirac operator provide 
an unambiguous determination of the topological charge ($Q_t$) of the gauge background, 
\bea\label{IndexTheorem}
Q_t \equiv \int d^4 x\, \rho_Q(x) = n_+ - n_- = \mathrm{index}(D_{\rm ov}), 
\eea
where $\rho_Q(x)$ is the topological charge density, and $n_\pm$ is
the number of zero modes of $D_{\rm ov}$ with $\pm$ chirality.
Then we can also obtain the topological susceptibility,
\bea
\chi_t = \int d^4 x \{ \rho_Q(x)\rho_Q(0) \} = \frac{\left<Q_t^2\right>}{\Omega}, 
\eea
which can be used to determine the low-energy constants (e.g., $ \Sigma $, $ F_\pi $) 
through the chiral perturbation theory (ChPT) \cite{Leutwyler:1992yt,Mao:2009sy}.
Moreover, the spontaneous chiral symmetry breaking is related to the spectrum 
of the low-lying modes near $ \lambda = 0 $, as depicted by the Banks-Casher relation \cite{Banks:1979yr}
\bea
\Sigma = - \langle \bar\psi \psi \rangle = \pi \rho(0), 
\eea
where $ \rho(0) $ is the density of eigenvalues near $ \lambda = 0 $.

In general, to project a subset of eigenmodes of a large sparse matrix
$A$, the standard procedure is to construct a Krylov subspace of $A$ from a
starting vector $r_0$,
\bea
\mathcal{K}(A,r_0) = \left\{ r_0,\, A r_0,\, A^2 r_0,\, \ldots,\, A^{m-1} r_0 \right\}.
\eea
Then an orthonormal basis of the $m$-dimensional subspace can be constructed,  
which are regarded as approximated eigenvectors of $A$, the {\em Ritz vectors}.

In general, the Arnoldi method \cite{Arnoldi,Sorensen}
is an efficent way to perform the ``implicit shifted QR iteration'' on the Krylov subspace 
during restarts, which has been implemented in the widely used package ARPACK. 
However, for Hermitian matrices, it is more advantageous to use
the Lanczos algorithm \cite{Lanczos}
to obtain the {\em orthonormal} basis in the Krylov subspace.
In lattice QCD, the objective is to obtain the low-lying eigenmodes of 
the positive-definite Hermitian matrix $ H^2 $, which has become a challenging large-scale computation. 
Thus it is crucial to develop highly efficient algorithms and paradigms, based on the 
Lanczos algorithm and its variants.   

This paper is organized as follows.
In section 2, we outline the underlying theme for the projecton of the
low-lying eigenmodes of the overlap Dirac operator in lattice QCD, 
and our implementation of the adaptive Thick-Restart Lanczos algorithms ($a$-TRLan). 
In section 3, we compare the performances of our code and the widely used ARPACK, 
and present our test results. Finally we conclude with some remarks.

%%%%%%%%%%%%%%%%%%%%%%%%%%%%%%%%%%%%%%%%%%%%%%%%%%%%%%%%%%%%%%%%%%%%%%%%
%%%%%%%%%%%%%%%%%%%%%%%%%%%%%%%%%%%%%%%%%%%%%%%%%%%%%%%%%%%%%%%%%%%%%%%%

\section{Projection of the low-lying eigenmodes of the overlap-Dirac operator}

In this section, we outline our scheme of projecting the low-lying eigenmodes of
the overlap-Dirac operator $D_{\mathrm{ov}}$ (\ref{Dovb}) for massless fermion.

The eigenvalues of $D_{\mathrm{ov}}$ are lying on a circle in
the complex plane with center at $ (m_0,0) $ and radius of length $ m_0 $, 
consisting of complex eigenmodes in conjugate pairs, 
and (for topologically nontrivial gauge background) real eigenmodes 
with eigenvalues at $0$ and $ 2 m_0 $ satisfying the chirality sum rule, 
$ n_+ - n_- + N_+ - N_- = 0 $ \cite{Chiu:1998bh}, 
where $ n_\pm ( N_\pm ) $ denote the number of eigenmodes at $0$ ($2 m_0 $) with $ \pm $ chirality. 
Empirically, the real eigenmodes always satisfy either 
($ n_+ = N_-$, $n_- = N_+ = 0 $) or ($ n_- = N_+ $, $n_+ = N_- = 0 $).
In other words, the zero modes only appear in either positive or negative chirality. 
Obviously, each eigenvector can be denoted by $ \ket{\theta} $, satisfying   
\bea
\label{eq:eigen-Dov}
D_{\mathrm{ov}}\ket{\theta} = \lambda(\theta)\ket{\theta},\hspace{2mm}
\lambda(\theta) = m_0 (1+e^{i\theta}), \hspace{2mm} \theta \in [0, 2 \pi).  
\eea
Using $ [D_{\mathrm{ov}} D_{\mathrm{ov}}^{\dagger}, \gamma_5]=0 $,
the eigenvector can be decomposed into positive and negative chiralities, 
\bea
\label{eq:eigen-Dov-pm}
S_{\pm}\ket{\theta}_{\pm} 
\equiv P_{\pm} S_{\mathrm{opt}}(H_w) P_{\pm} \ket{\theta}_{\pm}
= P_{\pm} H_w R_Z(H_w^2) P_{\pm} \ket{\theta}_{\pm}
= \pm \cos\theta\ket{\theta}_{\pm}, \quad 
\eea
where $P_{\pm} = (1 \pm \gamma_5)/2$, $\ket{\theta}_\pm = P_\pm \ket{\theta} $. 
Moreover, for complex eigenmodes, $ \ket{\theta}_+ $ and $ \ket{\theta}_- $ are related to each other through the relation  
\bea
\label{eq:DovEiv}
\ket{\theta}
= \frac{1}{i\sin\theta}(\gamma_5 S_{\mathrm{opt}} -e^{-i\theta})
  \ket{\theta}_\pm, \hspace{10mm} \theta \ne 0, \pi.
\eea
Thus our strategy of projecting the low-lying modes of $ D_{\rm ov} $ is as follows. 
First, we check whether the zero modes are in the positive or negative chirality sector.
Then we project the low-lying eigenmodes in the same chirality sector of the zeromodes 
(if there are no zeromodes, just pick either one of the chiralty sectors), 
and finally obtain the eigenvectors of the complex eigenmodes using (\ref{eq:DovEiv}). 
To compute the matrix-vector product $ R_Z(H_w^2) \ket{\theta}_\pm $ 
in (\ref{eq:eigen-Dov-pm}) and (\ref{eq:DovEiv}), we use multi-shift CG with low-mode preconditioning and  
the two-pass algorithm \cite{Neuberger:1998jk,Chiu:2003ub}. 
Therefore, we first project the low-lying eigenmodes of $H_w^2$,
before doing the low-mode projection of (\ref{eq:eigen-Dov-pm}). 
In both cases, we use the adaptive thick restart Lanczos algorithm ($a$-TRLan)
\cite{TRLan,a-TRLan} for the low-mode projection.  

In the following, we outline the basic ideas of $a$-TRLan. 
Consider a $N\times N$ Hermitian matrix $A$. After $m$ Lanczos iterations, it gives 
\bea
\label{LanczosFactor}
A Q_m = Q_m T_m + \beta_m q_{m+1} e^{\dagger}_m
\eea
where $T_m$ is a $m \times m$ tridiagonal matrix, and $Q_m$ is the unitary
matrix formed by the orthonormal vectors
$\{q_i, i=1,2,\ldots,m\}$ which are generated in the Lanczos iterations.
The last term is called the residual,
where $e_m$ is the unit vector with all elements zero except the $m$-th entry,
and $\beta_m$ is the norm of the residual.
After diagonalizing $T_m$, we obtain $m$ Ritz pairs, $ \{ (\lambda_i, y_i): \lambda_1 < \lambda_2 < \cdots < \lambda_m \} $.
Since the Ritz values first converge to the exterior eigenvalues of $A$, we
select $ k $ Ritz pairs from both ends for the restart, namely $ i=1,\cdots, l, u, \cdots, m $, where $ k = m + l-u + 1 $, 
and to perform re-orthogonalization if necessary.
Then the Lanczos iterations are restarted for $k+1, \cdots, m $.  
The same procedures (diagonalization, truncation, re-orthogonalization, and restart) 
are repeated until the desired number of Ritz pairs are converged. 

Theoretically, the subspace dimension $ m $ and the
number $ k $ of selected Ritz pairs for the restart, i.e., the triplet $ (l,u,m) $ can be
tuned for each restart to optimize the overall performance, i.e.,
the convergent rate versus the computational time of each restart,
based on the estimates from all previous iterations.
This leads to the $a$-TRLan algorithm \cite{a-TRLan}.
This amounts to maximizing the object function (i.e., the ratio of the convergence rate and the computation time) 
at each restart, 
\bea
\label{objFunc}
f(l,m,k) = \frac{\omega(m,k,\gamma)}{T(m,k)} \simeq \frac{ 2(m-k)\sqrt{\gamma} }{2 t_1(m-k)(k+m-1)+ 2 t_2 m k + t_3 (m-k)},   
\eea
where $ \omega $ measures the convergent rate of the smallest non-convergent Ritz value ($ \lambda_{l+1} $), 
which can be estimated as $ 2 (m-k) \sqrt{\gamma} $, with  
$ \gamma = (\lambda_{l+2} - \lambda_{l+1} )/(\lambda_{u-1} - \lambda_{l+2}) $, the effective gap ratio. 
Here $ t_1, t_2, t_3 $ are average computation times (measured from all previous iterations) 
for reorthogonalization, computing Ritz vectors, and matrix-vector product respectively. 

%%%%%%%%%%%%%%%%%%%%%%%%%%%%%%%%%
\begin{figure}[t!]
\begin{center}
\includegraphics[width=115mm,height=65mm]{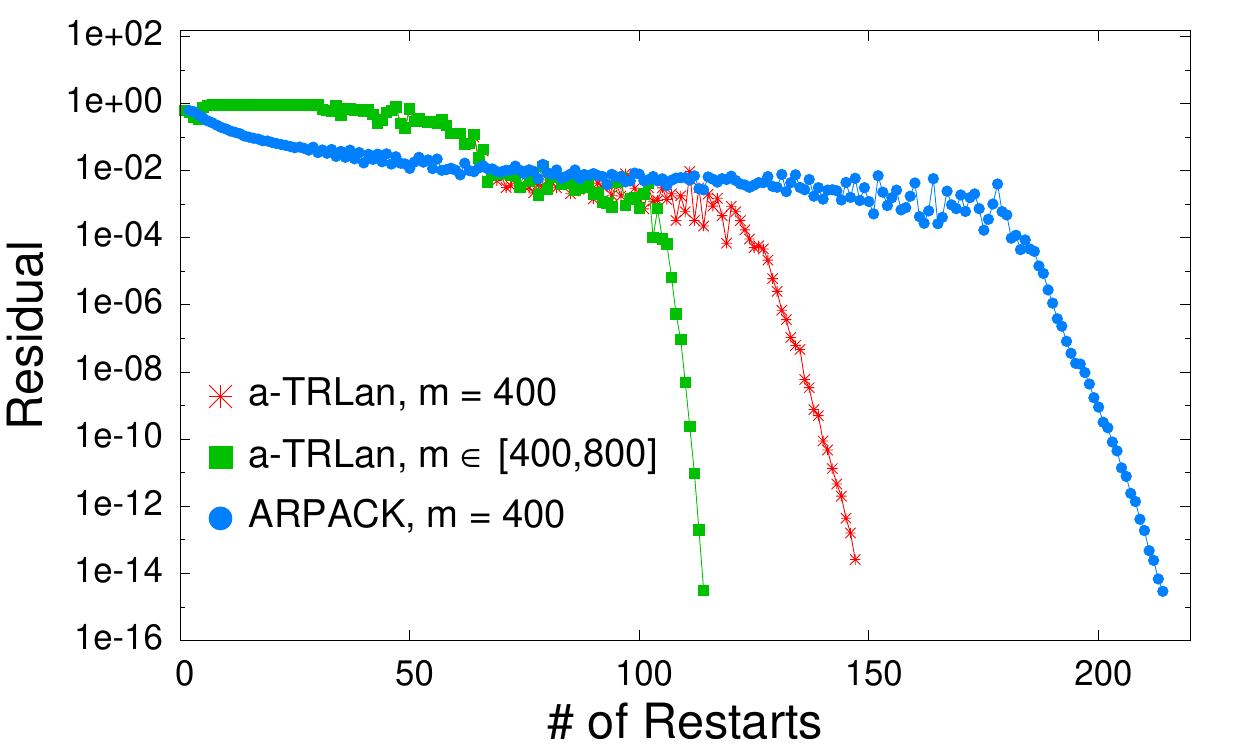}
\end{center}
\caption{\label{fig:residual}
The residual of the last (the 250-th) Ritz pair
of $H_w^2$ versus the number of restarts, for $a$-TRLan
with fixed $m$, $a$-TRLan with tunable $m$, and ARPACK respectively.
Here $m$ is the dimension of the Krylov subspace. For the $a$-TRLan cases
(the red and green points), the relaxation parameter $\nu$ is set to 0.5.
}
\end{figure}
%%%%%%%%%%%%%%%%%%%%%%%%%%%%%%%%%

According to our experience, taking $ u = m + 1 $ is always the best choice. 
Thus we restrict our discussions to the case $ k = l $. 
In our code, the variation of $ m $ is limited to an interval [$m_{\mathrm{min}}$, $m_{\mathrm{max}}$], 
with the initial $ m = m_{\rm min} $.
Assuming the number of converged Ritz pairs is $n_{\mathrm{conv}}$, 
then $ k $ is varied with the constraint $ n_{\rm conv} + 1 \le  k \le k_0 < m$,
where $k_0 = m + 1 - \nu (m - n_{\mathrm{conv}}) $, with $ \nu $ the relaxation (input) parameter.
At the $j$-th restart, after $ m_j $ Lanczos iterations, and the diagonalization of the tridiagonal matrix $ T_{m_j} $, 
the code searches for the optimal values of $ (m_{j+1},k_{j+1}) $ to 
maximize the object function (\ref{objFunc}) with the constraint $ k_{j+1} < m_{j} $, 
which are then used for the $ (j+1) $-th restart.

%%%%%%%%%%%%%%%%%%%%%%%%%%%%%%%%%%%%%%%%%%%%%%%%%%%%%%%%%%%%%%%%%%%%%%%%
%%%%%%%%%%%%%%%%%%%%%%%%%%%%%%%%%%%%%%%%%%%%%%%%%%%%%%%%%%%%%%%%%%%%%%%%

\section{Performance Tests}

%%%%%%%%%%%%%%%%%%%%%%%%%%%%%%%
\begin{figure}[th]
\begin{center}
\begin{tabular}{@{}c@{\ }c@{}}
\includegraphics[width=75mm]{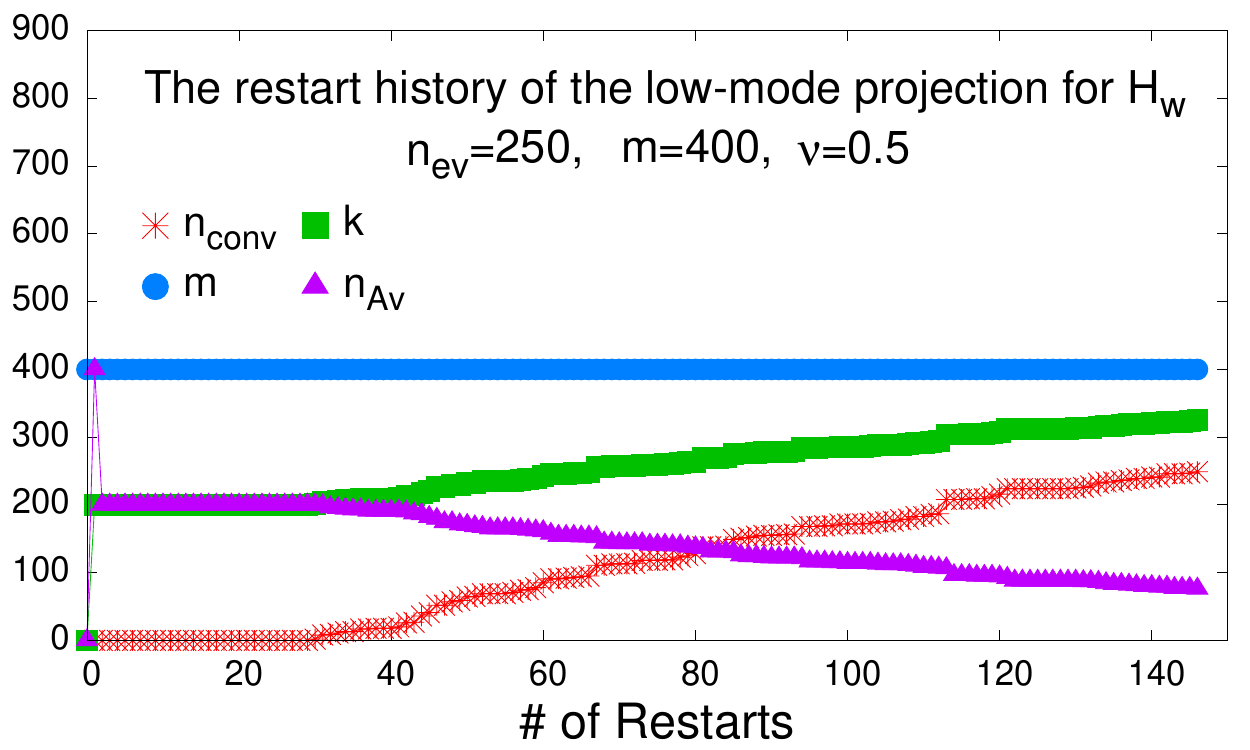} &
\includegraphics[width=75mm]{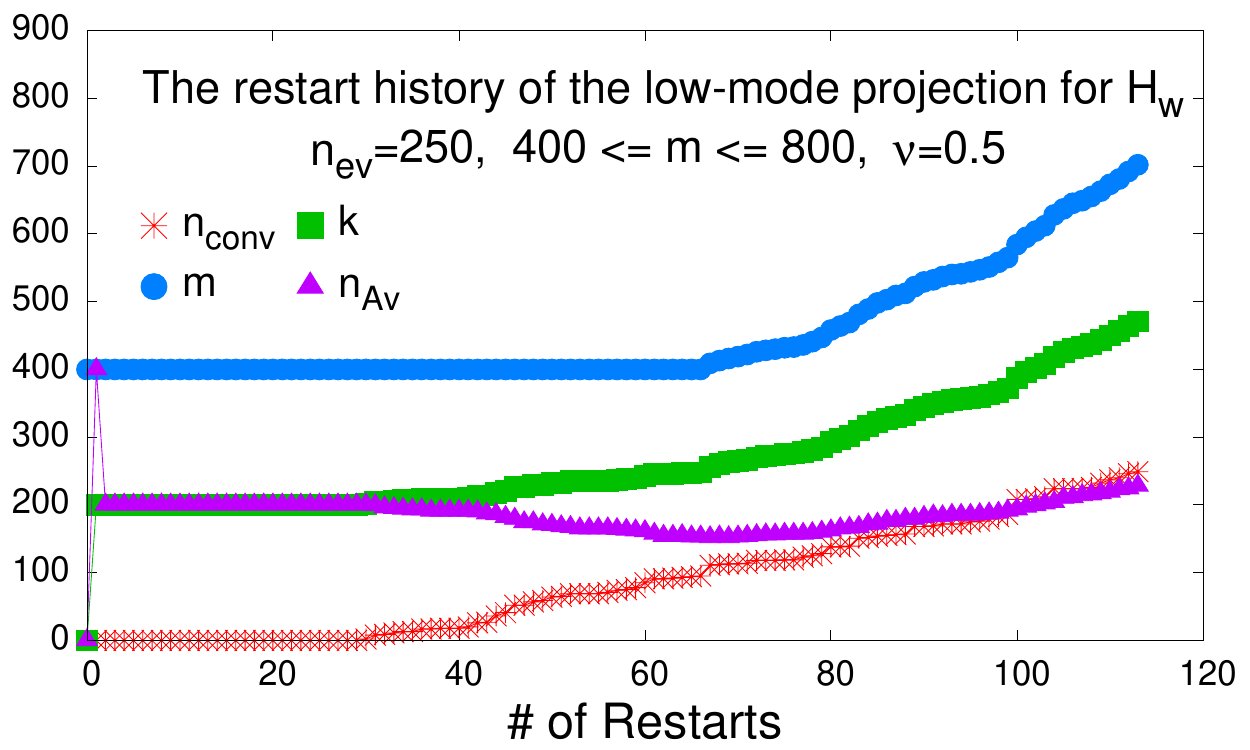} \\
(a) & (b)
\end{tabular}
\end{center}
\caption{\label{ProjHistory}
The restart history of projecting the low-modes of $ H_w^2 $ with the $a$-TRLan algorithm.
Here $n_{\mathrm{ev}}$ is the number of desired eigenmodes, $m$ is the
dimension of the Krylov subspace, $\nu$ is the relaxation factor.
$k$ is the dimension at restart,
$n_{\mathrm{conv}}$ is the number of converged Ritz pairs, and
$n_{Av}$ is the number of matrix-vector multiplication between each restart.
In (a) $m$ is fixed at 400, only $k$ is tunable
during the restarts. In (b) both $(m,k)$ are tunable during
the restarts, in which the range of $m$ is confined within $[400,800]$.}
\end{figure}
%%%%%%%%%%%%%%%%%%%%%%%%%%%%%%%

In this section, we perform some tests to compare 
the performances of our code and ARPACK.
Our testing platform is a Supermicro server with dual Intel Xeon E5530 (2.40 GHz) CPU and 24 GB memory.
There are totally 8 cores in the machine. Hence we run each of the tests 
with 8 threads in parallel.

In Fig. \ref{fig:residual}, we plot the convergence rate of the low-modes
of $H_w^2$, for a typical configuration in the gauge ensemble 
obtained by the dynamical simulation of 2-flavors lattice QCD 
with optimal domain-wall quarks \cite{Chiu:2002ir}
on the $16^3\times 32$ lattice at $\beta=5.95$ and $m_{sea} a = 0.01$ \cite{Chiu:2011bm}.
Here we compare the performances of three different schemes: 
(1) $a$-TRLan with fixed $m=400$, i.e., only tuning $ k $. 
(2) $a$-TRLan with tunable $(m,k)$ for $m\in [400,800]$. 
(3) ARPACK with fixed $m=400$. 
In each case, we project 250 low-lying eigenmodes of $H_w^2$, 
and plot the residual of the last (the 250-th) Ritz pair versus the number of restarts. 
The stopping criteira is $10^{-13}$ for the residual of any Ritz pair. 
For $a$-TRLan, the relaxation factor $ \nu $ is fixed at 0.5.
Here we see that $a$-TRLan takes much less restarts than ARPACK. 
Furthermore, $a$-TRLan with tunable $(m,k)$ outperforms that with fixed $m$. 
The computation time for each case is given in Table \ref{bench_Hw}.

In Fig. \ref{ProjHistory}, we plot the history of restart 
corresponding to the low-mode projection in
Fig. \ref{fig:residual}. In (a) the dimension of the Krylov subspace $m$
is fixed at 400, thus the optimization amounts to search for the optimal $k$ for the next restart. 
On the other hand, $m$ is not fixed in (b), thus the algorithm searches for the optimal values of $(m,k)$.
Here we see that the value of $k$ increases monotonically with respect to the number of converged Ritz pairs $n_{\mathrm{conv}}$.
In (a), since $m$ is fixed, the required Lanczos iterations after
each restart (i.e., $m-k-1$) gets fewer and fewer. On the other hand, in the case of (b), 
$m$ and $ k $ both increase monotonically with respect to $n_{\mathrm{conv}}$, the required Lanczos
iterations for each restart is almost the same, but the required number
of restarts is less than (a). Overall, (b) runs faster than (a) (see Table \ref{bench_Hw}).

%%%%%%%%%%%%%%%%%%%%%%%%%%%%%%%%%%%%
\begin{table}[th]
\begin{center}
\footnotesize
\begin{tabular}{c|r@{\ \ }r@{\ \ }r@{\ \ }c|r@{\ \ }r@{\ \ }r@{\ \ }c|r@{\ \ }r@{\ \ }r@{\ \ }c|r@{\ \ }r@{\ \ }c}
\hline
    & \multicolumn{4}{|c}{$\nu=0.7$} &
      \multicolumn{4}{|c}{$\nu=0.6$} &
      \multicolumn{4}{|c}{$\nu=0.5$} &
      \multicolumn{3}{|c}{ARPACK} \\
\cline{2-16}
$m$ & $m_M$ & $n_R$ & $n_{Av}$ & time &
      $m_M$ & $n_R$ & $n_{Av}$ & time &
      $m_M$ & $n_R$ & $n_{Av}$ & time &
              $n_R$ & $n_{Av}$ & time \\
\hline
300 & 300   & 361 & 33196 & 40095 &
      300   & 346 & 28137 & 36799 &
      300   & 361 & 33196 & 40095 &
             1710 & 47728 & 163010 \\
400 & 400   & 128 & 24433 & 30907 &
      400   & 121 & 21502 & 26343 &
      400   & 146 & 21569 & 29168 &
              220 & 22255 & 56702 \\
500 & 500   &  75 & 21466 & 30928 &
      500   &  88 & 21526 & 30197 &
      500   &  75 & 21466 & 31286 &
              120 & 21571 & 57227 \\
600 & 600   &  59 & 21445 & 32743 &
      600   &  69 & 21468 & 33949 &
      600   &  59 & 21445 & 32426 &
               78 & 21262 & 59260 \\
$[300,800]$ & 412 & 243 & 31419 & 30791 &
              458 & 203 & 26143 & 26907 &
              565 & 186 & 24064 & 24581 &
              n/a &  n/a  &  n/a  \\
$[400,800]$ & 430 & 126 & 24387 & 21802 &
              453 & 118 & 21510 & 25858 &
              558 & 132 & 21546 & 23905 &
                    n/a &  n/a  &  n/a  \\
$[500,800]$ & 500 &  75 & 21466 & 30994 &
              500 &  88 & 21526 & 30224 &
              546 & 104 & 21478 & 29620 &
                    n/a &  n/a  &  n/a  \\
$[600,800]$ & 600 &  59 & 21445 & 32778 &
              600 &  69 & 21468 & 34016 &
              600 &  83 & 21499 & 32513 &
                    n/a &  n/a  &  n/a  \\
\hline
\end{tabular}
\end{center}
\caption{\label{bench_Hw}
Comparing performances of various schemes of $a$-TRLan and ARPACK, for the projection of low-modes of $ H_w^2 $.
For $a$-TRLan, both fixed $ m $ and tunable $m$ are considered 
for various settings of the relaxation factor $\nu=0.7$, 0.6, and 0.5.
Here $m_M$ is the maximum dimension
of the Krylov subspace in the projection, $n_R$ is
the number of restarts to complete the job, $n_{Av}$ is the total number
of matrix-vector multiplication, and the total computation time is in unit of second.}
\end{table}
%%%%%%%%%%%%%%%%%%%%%%%%%%%%%%%%%%%%

In Table \ref{bench_Hw}, we compare the performance of projecting
250 low-lying eigenmodes of $H_w^2$ for the same gauge configuration
presented in Fig. \ref{fig:residual}. For the $a$-TRLan algorithm,
we also compare different settings of $\nu$ and $m$,
with both cae of fixed $ m $ and tunable $ m $. Here we show how
the setting of $m$ and $\nu$ affects the performance. For the cases with
fixed $m$, it seems that setting $m$ at around 400 or 500 can attend
a reasonably good performance for all the values of $\nu$, and setting
$\nu=0.6$ seems to be the best choice. Now we compare the results of
fixing $m=m_{\mathrm{min}}$ and tunable $m\in [m_{\mathrm{min}},800]$,
where $m_{\mathrm{min}}=300$, 400, 500, and 600 are the lower
bounds for the tuning $m$. We see that in some cases the maximum values of
$m$ (i.e., $m_M$) during the projection may saturate. For example,
at $\nu=0.7$, setting the lower bound $m_{\mathrm{min}}$ to 500 or 600,
$m_M$ (for the tunable cases) remains constant during the entire projection. 
Thus, the restart histories of both fixed $m=m_{\mathrm{min}}$ and tunable
$m\in [ m_{\mathrm{min}}, 800 ]$ are exactly identical. 
On the other hand, if $m_M$ is not saturated, then those with tunable $m$ outperform 
their counterparts with fixed $m$.
In the last three columns of Table \ref{bench_Hw}, we list the performance
of ARPACK for the same projection task.
Here neither $m$ nor $k$ is tunable in ARPACK. Obviously, 
any scheme of $a$-TRLan in Table \ref{bench_Hw} outperforms ARPACK.

%%%%%%%%%%%%%%%%%%%%%%%%%%%%%%%%%%%%
\begin{table}[th]
\begin{center}
\footnotesize
\begin{tabular}{c|r@{\ \ }r@{\ \ }r@{\ \ }c|r@{\ \ }r@{\ \ }r@{\ \ }c|r@{\ \ }r@{\ \ }r@{\ \ }c|r@{\ \ }r@{\ \ }c}
\hline
    & \multicolumn{4}{|c}{$\nu=0.6$} &
      \multicolumn{4}{|c}{$\nu=0.5$} &
      \multicolumn{4}{|c}{$\nu=0.4$} &
      \multicolumn{3}{|c}{ARPACK} \\
\cline{2-16}
$m$ & $m_M$ & $n_R$ & $n_{Av}$ & time &
      $m_M$ & $n_R$ & $n_{Av}$ & time &
      $m_M$ & $n_R$ & $n_{Av}$ & time &
              $n_R$ & $n_{Av}$ & time \\
\hline
300 & 300   & 13 & 2002 & 65657 &
      300   & 15 & 1967 & 64754 &
      300   & 19 & 1968 & 64744 &
              27 & 1986 & 77613   \\
400 & 400   &  8 & 2029 & 66827 &
      400   &  9 & 1961 & 64619 &
      400   & 12 & 2010 & 66249 &
              12 & 1966 & 78014   \\
500 & 500   &  6 & 2110 & 69561 &
      500   &  7 & 2072 & 68479 &
      500   &  6 & 1950 & 64730 &
               8 & 2082 & 83368   \\
600 & 600   &  4 & 2003 & 66130 &
      600   &  5 & 2038 & 67498 &
      600   &  6 & 1988 & 65884 &
               5 & 1996 & 79350   \\
$[300,800]$ & 524 & 9 & 2126 & 69799 &
              597 & 8 & 2017 & 66454 &
              759 & 7 & 2018 & 66580 &
              n/a &  n/a  &  n/a  \\
$[400,800]$ & 460 & 7 & 1979 & 65134 &
              535 & 7 & 1951 & 64260 &
              685 & 7 & 2030 & 67240 &
                    n/a &  n/a  &  n/a  \\
$[500,800]$ & 532 & 6 & 2147 & 70828 &
              589 & 6 & 2026 & 66835 &
              651 & 6 & 1950 & 64729 &
                    n/a &  n/a  &  n/a  \\
$[600,800]$ & 600 & 4 & 2003 & 66238 &
              627 & 5 & 2065 & 68286 &
              817 & 5 & 2131 & 70770 &
                    n/a &  n/a  &  n/a  \\
\hline
\end{tabular}
\end{center}
\caption{\label{bench_Dov}
The performance comparison of low-mode projection for $S_+$
using the $a$-TRLan algorithm with fixed and tunable $m$, and
for various settings of the relaxation factor $\nu=0.6$, 0.5, 0.4,
together with the performance of ARPACK. Here $m_M$ is the maximum dimension
of the Krylov subspace during the projection, $n_R$ is
the number of restarts to complete the job, $n_{Av}$ is the total number
of matrix-vector multiplication, and the running time is in unit of seconds.}
\end{table}
%%%%%%%%%%%%%%%%%%%%%%%%%%%%%%%%%%%%

Now, with 250 low-modes of $ H_w^2 $, we proceed to project the 200 low-modes of the overlap Dirac operator.
For the matrix-vector product $ R_Z(H_w^2) \ket{\theta}_\pm $ in (\ref{eq:eigen-Dov-pm}) and (\ref{eq:DovEiv}), 
we use multi-shift CG with low-mode preconditioning and the two-pass algorithm \cite{Neuberger:1998jk,Chiu:2003ub}. 
In Table \ref{bench_Dov}, we present our test results of low-mode projection of $S_+$ (\ref{eq:eigen-Dov-pm})
for the same gauge configuration. 
Note that $S_\pm \cdot v $ is a complicated matrix-vector operation since
it involves two inner CG loops (for the two-pass algorithm) \cite{Neuberger:1998jk,Chiu:2003ub}, 
which takes much longer time than $H_w^2 \cdot v $. 
For this configuration, the ratio of the computation times of these two matrix-vector multiplications is  
$$
\frac{T_{Av}(S_+)}{T_{Av}(H_w^2)} \simeq 1120. 
$$
Thus, for the low-mode projection of $S_\pm$, the computation time 
is dominated by the total number of matrix-vector multiplication $n_{Av}$. 
Consequently, $a$-TRLan with tunable $m$ is not necessarily faster than
that with fixed $m$.
Even though the object function is supposed to predict the optimal values of $(m,k)$,  
it does not guarantee to yield the smallest value of $n_{Av}$ which 
dominates the cost of the computation.
This issue requires further studies which are beyond the scope of the present paper.
From Table \ref{bench_Dov}, setting $\nu=0.5$ and $m=400$ seems to give
the best performance. 
%Moreover, turning off the option of tuning $m$ also saves a lot of memory. 
From the last 3 columns in Table \ref{bench_Dov}, we see that 
the performance of $a$-TRLan is only 20\% to 30\% better than ARPACK, 
unlike the case of $ H_w^2 $ (see Table \ref{bench_Hw}). 
We believe that there is still room for improving the performance of   
the low-mode projection of $S_\pm$ with $a$-TRLan. 

At this point, it is instructive to compare the overall performances
of our $a$-TRLan code and ARPACK, for the projection of 200 low-lying 
modes of the overlap Dirac operator, including the time in the  
projection of 250 low-modes of $ H_w^2 $. Taking into account of the 
time (6550 seconds) used in evaluating (\ref{eq:DovEiv}), our results 
in Tables \ref{bench_Hw} and \ref{bench_Dov}
suggest that our $a$-TRLan code is about 1.5 times faster than ARPACK.      
Moreover, if one is only interested in the projection of low-modes of the 
Wilson (clover) quark matrix, our results in Table \ref{bench_Hw} 
suggest that $a$-TRLan is more than 2 times faster than ARPACK.

%%%%%%%%%%%%%%%%%%%%%%%%%%%%%%%%%%%%%%%%%%%%%%%%%%%%%%%%%%%%%%%%%%%%%%%%
%%%%%%%%%%%%%%%%%%%%%%%%%%%%%%%%%%%%%%%%%%%%%%%%%%%%%%%%%%%%%%%%%%%%%%%%

\section{Concluding remarks}

In this work, we have implemented the $a$-TRLan algorithm to project the
low-lying eigenmodes of the Dirac operators $H_w$ and $D_{\mathrm{ov}}$.
Our code searches for the optimal values of $(m,k)$ at each restart to maximize 
the object function (\ref{objFunc}) which is the ratio of the convergence rate 
(of the smallest non-convergent Ritz value) and the computation time. 
For the low-mode projection of $H_w^2$, $a$-TRLan works very well 
and outpeforms ARPACK significantly.  
However, for the low-mode projection of $S_\pm$, $a$-TRLan with fixed $m$
is only about 20\%-30\% faster than ARPACK. Moreover, $a$-TRLan with 
tunable $ (m,k) $ does not perform better than that with fixed $ m $.
To clarify this issue requires further studies which are now in progress.  

To summarize, we find that $a$-TRLan is an efficient algorithm for the projection 
of low-modes of the quark matrix in lattice QCD. 
For the Wilson (clover) quark matrix, $a$-TRLan is more than 2 times faster than ARPACK.
For lattice QCD with exact chiral symmetry, the projection of the low-modes of the quark matrix 
includes two different projections (i.e., those of $ H_w^2 $ and $ S_{\pm} $).
Currently, our $a$-TRLan code is about 1.5 times faster than ARPACK.

\begin{acknowledgments}
  TWC would like to thank Horst Simon, Sherry Li, and Kesheng Wu for their hospitality during his visit 
  to Lawrence Berkeley National Laboratory, and also for discussions on $a$-TRLan. 
  This work is supported in part by the Ministry of Science and Technology
  (Nos.~NSC102-2112-M-002-019-MY3,~NSC102-2112-M-001-011) and NTU-CQSE (No.~103R891404).
  We also thank NCHC for providing facilities to perform part of our calculations.
\end{acknowledgments}

%%%%%%%%%%%%%%%%%%%%%%%%%%%%%%%%%%%%%%%%%%%%%%%%%%%%%%%%%%%%%%%%%%%%%%%%
%%%%%%%%%%%%%%%%%%%%%%%%%%%%%%%%%%%%%%%%%%%%%%%%%%%%%%%%%%%%%%%%%%%%%%%%

\end{document}